\documentclass[vecphys]{svmult}
\usepackage{graphicx}  
     
\begin{document}

\title*{Gravitationally Lensed QSOs: Optical Monitoring with the EOCA and the Liverpool Telescope (LT)}
\titlerunning{GLQSOs: Optical Monitoring with the EOCA and the LT}
\author{L. J. Goicoechea, A. Ull\'an\inst{1}\, J. E. Ovaldsen\inst{2}\, E. Koptelova\inst{3}\, V. N.
Shalyapin\inst{4}\ and R. Gil-Merino\inst{5}}
\authorrunning{Goicoechea et al.} 
\institute{Universidad de Cantabria, Spain
\texttt{goicol@unican.es, aurora.ullan@postgrado.unican.es}
\and University of Oslo, Norway \texttt{j.e.ovaldsen@astro.uio.no}
\and Moscow University, Russia \texttt{koptelova@xray.sai.msu.ru}
\and National Academy of Sciences, Ukraine \texttt{vshal@ire.kharkov.ua}
\and University of Sydney, Australia \texttt{rodrigo@physics.usyd.edu.au}}

\maketitle

\section{The aim}
\label{sec:1}
The aim of this contribution is to present the two first phases of the optical monitoring programme 
of the Gravitational Lenses group at the Universidad de Cantabria (GLUC, http://grupos.unican.es/glendama/). 
In an initial stage (2003 March-June), the Estaci\'on de Observaci\'on de Calar Alto (EOCA) was used 
to obtain $VR$ frames of SBS 0909+532 and QSO 0957+561. These observations in 2003 led to accurate fluxes 
of the two components of both double QSOs, which are being compared and complemented with data from other 
1-1.5 m telescopes located in the North Hemisphere: Fred Lawrence Whipple Observatory (USA), Maidanak 
Observatory (Uzbekistan) and Wise Observatory (Israel). On the other hand, the GLUC started the second 
phase of its monitoring programme in 2005 January. In this second phase, they are using the 2 m fully 
robotic Liverpool Telescope (LT). The key idea is the two-band photometric follow-up of four lensed QSOs 
with different main lensing galaxies: SBS 0909+532 (elliptical), QSO 0957+561 (giant cD), B1600+434 
(edge-on spiral) and QSO 2237+0305 (face-on spiral). Thus, the light rays associated with the components of 
the four gravitational mirages cross different galaxy environments, and the corresponding light curves 
could unveil the content of these environments. While SBS 0909+532 and QSO 0957+561 are the targets for the
two first years with the LT (2005-2006), the rest of targets (B1600+434 and QSO 2237+0305) will be 
monitored starting from 2007.

The photometry and the analysis of data are conducted by the GLUC members and collaborators in several 
non-Spanish institutes, and we focus on the most relevant measurements, i.e., detection of intrinsic events
(caused by the sources), time delay and flux ratio estimates, quantitative study of the structure and 
chromaticity of the intrinsic signals, determination of chromatic (between two optical filters) delays, and 
detection of extrinsic fluctuations (due to intervening objects in the involved galaxies, e.g., 
microlenses or dusty clouds). These measurements are used to tackle three $hot$ astrophysical subjects: 
current expansion rate of the Universe, structure of the main lensing galaxies and nature of the sources 
(QSOs). The discovery of variable field stars or supernova explosions might be a bonus of the programme.

\section{SBS 0909+532}
\label{sec:2}
SBS 0909+532 consists of two components (double QSO) separated by about 1 arcsec \cite{Kochetal, Oscetal, 
Lehetal, Lubetal}. The lensing elliptical galaxy has a large effective radius with a correspondingly low 
surface brightness \cite{Lehetal}. In Fig. 1 (top panel) we show a Maidanak subframe in the $R$ band, where 
there are two close quasar components, but the very faint galaxy is not apparent. Although the presence of a
faint galaxy has a positive aspect: simple photometric model with only two close point-like sources 
(we can avoid additional complications arising from the use of a galaxy profile), the faint and extended 
lenses are not a good business for cosmological studies. The relative astrometry of the quasar components 
(with respect to the centre of the lens galaxy) is not accurate, and this fact plays a crucial role in the 
estimation of the Hubble constant or the surface density of the lens (see below).

During the {\it Phase I monitoring (EOCA/2003)}, the set of optical frames cover the period between 2003 March 4 
and June 2. Exposures in the $V$ and $R$ Johnson-Cousins filters were taken every night when clear. We have 
checked (via Wise data) the reliability of the EOCA differential photometry between widely separate and 
neighbouring stars (see the six field stars in the bottom panel of Fig. 1), and we have concluded that only the 
relative fluxes for neighbouring objects seem to be reliable. Therefore, as the {\it photometric ruler} 
(empirical PSF to compare with) is the normalised 2D profile of the "a" star and there is clear evidence of 
variability of the "c" star, measurements of the brightness of both quasar components (A and B) are always made 
with respect to the brightness of the nearby non-variable star "b". The $R$-band EOCA light curves are 
complemented with $R$-band fluxes from Maidanak frames, so the global monitoring period and the mean sampling 
rate (in the $R$ band) are 4 months ($\sim$ 120 days) and one point each 6 days, respectively. These are the 
first resolved brightness records of SBS 0909+532 \cite{Ulletal1}.

\begin{figure}
\centering
\includegraphics[angle=-90,width=4cm]{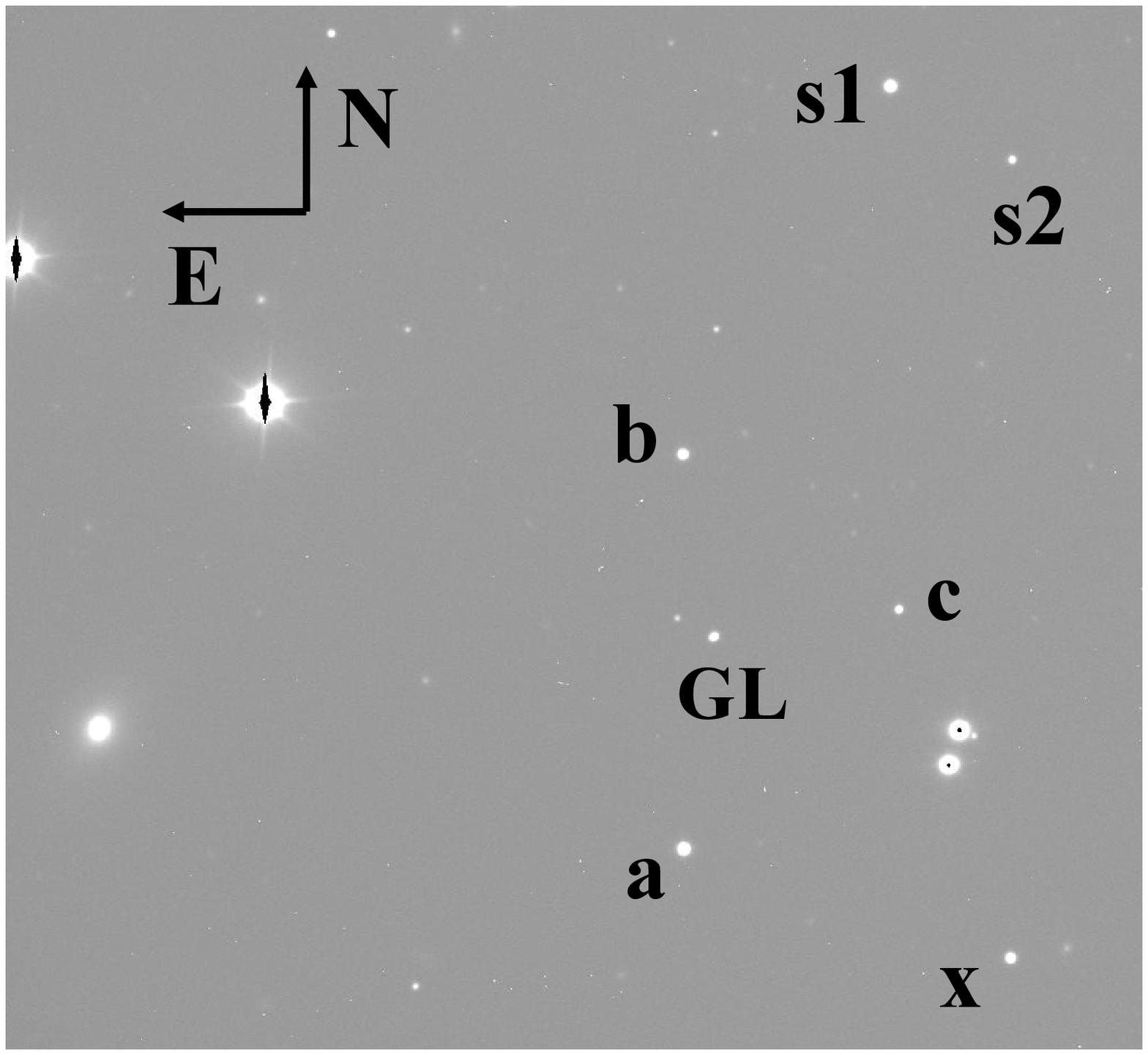}
\includegraphics[angle=0,width=4cm]{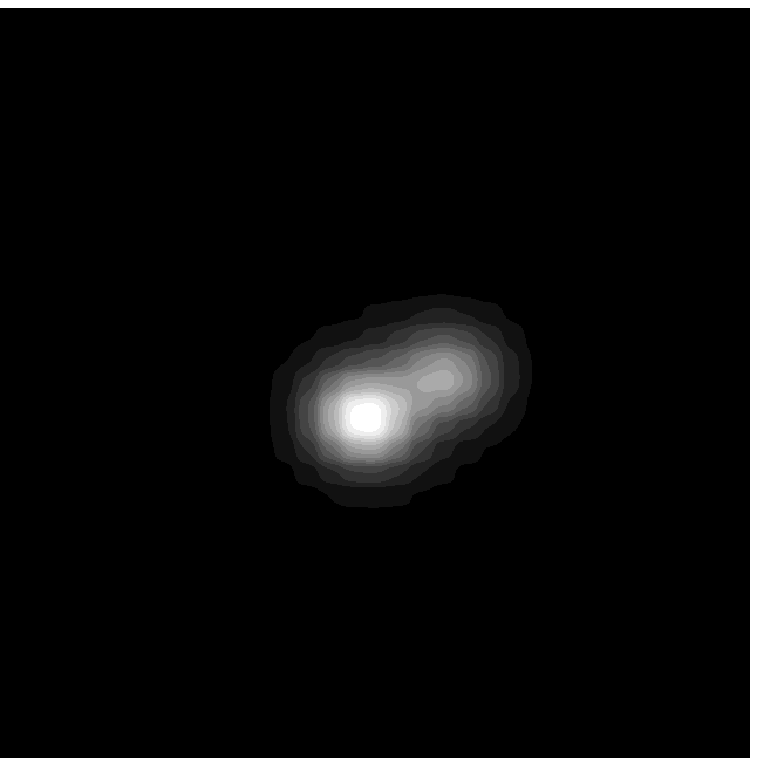}
\caption{{\it Top panel}: Maidanak image of SBS 0909+532 (about 8'' on a side). We see the two quasar components 
A (the brightest object) and B (the faintest object), i.e., the GL in the bottom panel. {\it Bottom panel}: EOCA 
image of the SBS 0909+532 field ($\sim$ 7' $\times$ 7'). This field contains the gravitationally lensed quasar 
(GL) and six bright and non-saturated  stars (a-c, s1-s2 and x).}
\label{fig:1}       
\end{figure}

Our data do not show any evidence for extrinsic fluctuations, and they can be interpreted as intrinsic signals. 
Moreover, the direct AB cross-correlation leads to very accurate 1$\sigma$ determinations of the time delay 
($\Delta \tau_{BA} = - 45 \pm 1$ days, where the sign "$-$" means that the intrinsic signal is observed first in 
B and later in A) and the flux ratio in the $R$ band. Thus, the EOCA + Maidanak light curves rule out a delay 
close to three months, which has been favoured in a recent prediction by the COSMOGRAIL collaboration 
\cite{Sahetal}. This longer delay is viable only if a very particular extrinsic variability is considered. 
However, the new analysis has one weak point that we want to comment on here. There is a relatively poor overlap 
between the A and B records, when the A light curve is shifted by the best solutions of the time delay and the 
flux ratio, so we are trying to get a confirmation of the delay and the absence of extrinsic signal. The LT is
being used for this purpose. On the other hand, the inaccurate position of the lensing galaxy (see above) does 
not permit to accurately measure the cosmic expansion rate ($H_0$) and the mean surface density of the lens
($<\kappa>$) \cite{Refs, Kocha}. Using the concordance cosmological model, the redshifts of the deflector and 
the source, basic astrometry of the system and the measured delay, our 1$\sigma$ contraints are $H_0$ = 82 $\pm$ 
41 km s$^{-1}$ Mpc$^{-1}$ (isothermal profile) and $1 - <\kappa>$ = 0.43 $\pm$ 0.21 ($H_0$ = 70 km s$^{-1}$ 
Mpc$^{-1}$) \cite{Ulletal2}. In order to obtain new accurate astrometry of SBS 0909+532, we are applying for 
observation time in Cycle 16 of the Hubble Space Telescope.

\begin{figure}
\centering
\includegraphics[angle=-90,width=8cm]{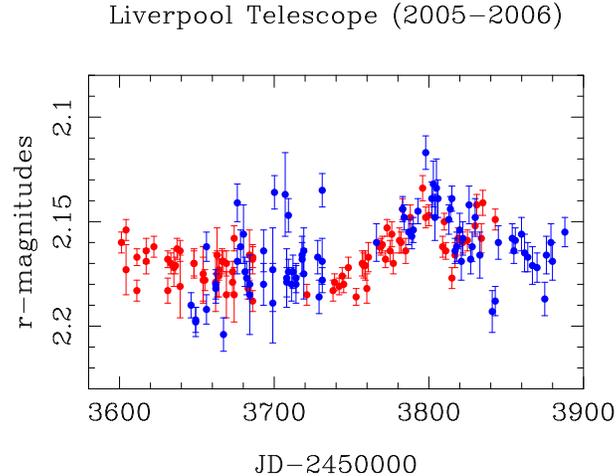}
\caption{Preliminary $r$-band comparison between the shifted SBS 0909+532A light curve (red circles) and the SBS 
0909+532B light curve (blue circles). In order to shift the A record, we use the EOCA + Maidanak delay, $\Delta 
\tau_{BA}$ = - 45 days, and a magnitude offset $\Delta m_{BA}$ = 0.65 mag.}
\label{fig:2}       
\end{figure}

The {\it Phase II monitoring (LT/2005-2006)} is more ambitious than the previous one. We take exposures in the
$g$ and $r$ Sloan filters (RATCAM optical CCD camera), and the robotic procedure is optimised to get frames all 
nights when SBS 0909+532 is visible (no occultation) if {\it ideal} conditions happen (no rain, no clouds, no 
technical problems, no fires...). From 2005 October to 2006 June we achieved a good sampling, so we focus in the 
preliminary $r$-band light curves of this double QSO in that period. In relation to the EOCA + Maidanak campaign 
in 2003 ($R$-band), the new 2005 October-2006 June season with the LT ($r$-band) has a better time coverage and 
resolution. Now we have a global monitoring period and a mean sampling rate of 8 months ($\sim$ 240 days) and 
one point each 3 days, respectively. In Fig. 2 we use the EOCA + Maidanak delay (see above) and a magnitude 
offset $\Delta m_{BA}$ = 0.65 mag to compare the B fluxes (blue circles) and the shifted fluxes of A (red 
circles). There is a reasonable agreement between the two components in Fig. 2, and these preliminary results 
encourage us. After to set a final selection criterium and include all the available fluxes, we hope to obtain
accurate measurements (variability, time delay, flux ratio, etc) as well as important astrophysical information  
\cite{Shaetal}. 

\section{QSO 0957+561}
\label{sec:3}
The double quasar QSO 0957+561 was discovered 27 years ago in a radio survey \cite{Waletal}. At optical wavelengths,
the main lens galaxy in the system appears as a relatively bright source close to the image B. This giant cD galaxy 
is part of a cluster of galaxies that also contributes to the lensing \cite{Stoc, Garetal}. The angular separation 
between B image and the centre of the cD galaxy is only of $\sim$ 1'', whereas the angular separation between A 
image and the galaxy is about five times larger \cite{Beretal}. In spite of almost 30 years of optical monitoring, 
there are several points of view about the origin of the variability in the QSO components. For example, while some 
people think that the records contain both long ($\sim$ years) and short ($\sim$ 100 days) timescale microlensing 
(extrinsic) events \cite{Schild}, other people have a very different opinion: there are no clear evidences for 
microlensing variability \cite{Giletal}. From previous $g$-band light curves of the system, it was also inferred 
the existence of two different delays (associated with 100-day intrinsic events of amplitude $\sim$ 100 mmag), 
$\Delta \tau_{BA} = 417.0 \pm 0.6$ days and $\Delta \tau_{BA} = 432.0 \pm 1.9$ days (1$\sigma$) \cite{Goico}. The 
time delay difference (15 $\pm$ 2 days) leads to a strong constraint for the separation of the two flares (in the 
far QSO) that cause the $g$-band intrinsic events: $\delta r \geq$ 300 pc (concordance cosmology). 

\begin{figure}
\centering
\includegraphics[angle=-90,width=8cm]{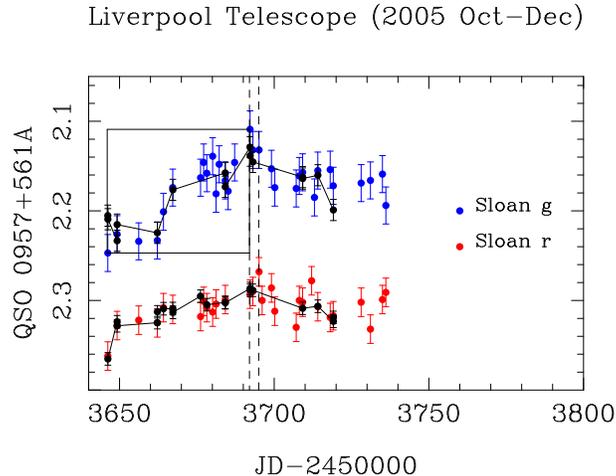}
\caption{LT light curves of QSO 0957+561A: deconvolution technique (blue/red points) and PSF fitting method (black 
points/lines). We use arbitrary zero points in the $gr$ flux (magnitude) scales.}
\label{fig:3}       
\end{figure}

The $VR$ frames taken in 2003 March-June ({\it Phase I monitoring}) are consistent with flat light curves for the
B component. However, we detect moderate events of about 50 mmag (amplitude) in the records of A (lasting 10-30 
days). In order to fully describe these $VR$ fluctuations, we looked for additional $VR$ images, which were taken
in the Fred Lawrence Whipple Observatory (FLWO) and the Wise Observatory. Up to now we have only analysed the EOCA + 
FLWO records in the $V$ and $R$ bands. We do not find clear evidences in favour of chromaticity, but a proper 
interpretation of the events will require more effort and data (Wise?). Complementary data could help to fill the 
gaps and trace the moderate fluctuations with better accuracy \cite{Ulletal2}.   

Finally, just one year ago (2005 October-November) the LT catched a prominent gradient in the $g$-band brightness 
of the A component ({\it Phase II monitoring}), so a similar gradient will be very probably seen in the $g$-band
light curve of B in 2006 December-2007 January ($\Delta \tau_{BA} \sim$ 14 months). A detailed reduction of frames 
from two different techniques (deconvolution and PSF fitting) indicates the existence of a prominent fluctuation in
the $g$-band record of A, which seems to be chromatic, with a smaller fluctuation in the $r$ band (see Fig. 3). 
This exciting discovery will probably permit to measure a new high-quality $g$-band delay, the structure and 
chromaticity of the intrinsic signal, etc.

{\it Acknowledgements}. We are indebted to J. Alcolea (OAN, Spain), B. Artamonov (Moscow University, Russia), E. 
Battaner (Universidad de Granada, Spain), D. Maoz (Tel-Aviv University, Israel), C. Moss (LT Support Astronomer),
E. Ofek (Tel-Aviv University, Israel), R. Schild (CfA, USA), R. Stabel (University of Oslo, Norway), A.P. Zheleznyak 
(Kharkov National University, Ukraine) and other colleagues for different kinds of support.

\end{document}